\documentclass[12pt]{book}
\usepackage{amsmath}
\usepackage{array}
\usepackage[doublespacing]{setspace}

\input{tcilatex}
\renewcommand{\baselinestretch}{2}

\begin{document}

\renewcommand{\baselinestretch}{1.2} %\lhead[\fancyplain{} \leftmark]{}
%\chead[]{}
%\rhead[]{\fancyplain{}\rightmark}
%\cfoot{}
%\headrulewidth=0pt
\markright{
} 
\markboth{\hfill{\footnotesize\rm Cheng-Yuan Liou and  Bruce R. Musicus
}\hfill}
{\hfill {\footnotesize\rm Cross Entropy Approx of Structured Covariance Matrices} \hfill}
\renewcommand{\thefootnote}{} $\ $

\fontsize{10.95}{14pt plus.8pt minus .6pt}\selectfont\vspace{0.812pc} %
\centerline{\large\bf Cross Entropy Approximation of Structured} \vspace{2pt}
\centerline{\large\bf Covariance Matrices} \vspace{0.4cm} %
\centerline{Cheng-Yuan Liou$^{1}$ and Bruce R. Musicus$^{2}$} \vspace{0.4cm} %
\centerline{\it  } \vspace{0.55cm} \fontsize{9}{11.5pt plus.8pt minus .6pt}%
\selectfont

\begin{quotation}
\noindent \textit{Abstract:} \ We apply two variations of the principle of
Minimum Cross Entropy (the Kullback information measure) to fit
parameterized probability density models to observed data densities. For an
array beamforming problem with $P$ incident narrowband point sources, $N>P$
sensors, and colored noise, both approaches yield eigenvector fitting
methods similar to that of the MUSIC algorithm[1]. Furthermore, the
corresponding cross-entropies are related to the MDL model order selection
criterion[2].

\vspace{9pt} \noindent \textit{Key words and phrases:} Array Beamforming,
Eigenvector methods, Kullback Information Measure, Minimum Cross Entropy,
Stochastic Estimation, Structured Covariance
\end{quotation}

\fontsize{10.95}{14pt plus.8pt minus .6pt}\selectfont\noindent \textbf{1.
Introduction}

\bigskip Many existing high resolution methods for spectral analysis and for
optimal beamforming utilize covariance matrices estimated from observed
data. Often, an underlying structure for the covariance matrix is known in
advance, and our goal is to estimate the covariance matrix with this
structure which best fits the observed data. Previous literature has
suggested a variety of methods of optimally estimating structured covariance
matrices from data[3,4,5]. In this paper, we will apply the minimum cross
entropy (CE)[6,7] and minimum reverse cross-entropy (RCE)[6] principles to
estimate the covariance matrix. These principles have proved to be quite
powerful in a wide variety of signal processing applications[8,9] and have
been justified as being "optimal" under suitable assumptions. In section 2,
we apply the CE and RCE procedures to the problem of estimating structured
covariance matrices, and in section 3 we demonstrate the utility of the idea
for a beamforming application.

.

\noindent \textbf{2. Problem Statement}

Let \underline{$x$} be an N-dimensional real or complex random vector.
Assume that a Gaussian probability density for \underline{$x$} is either
known a prior, or has been estimated by some procedure from observed
data:\bigskip 
\begin{equation}
p(\underline{x})=N(\underline{m},R)
\end{equation}%
where $\underline{m}$ is the expected value of \underline{$x$}, and $R$ is
the covariance matrix, $R=E[\underline{xx}^{H}]$, and where \underline{$x$}$%
^{H}$ is the Hermitian (complex conjugate transpose) of \underline{$x$}.
Suppose we wish to approximate this $p(\underline{x})$ with a parameterized
probability density function (PDF):%
\begin{equation}
q_{\theta }(\underline{x})=N(\underline{m}_{\theta },R_{\theta })
\end{equation}%
where \underline{$\theta $} denotes the unknown parameters $\theta $ in the
model $q_{\theta }(\underline{x})\ $which are to be estimated. Conceptually,
we wish to choose $\theta $ to make $q_{\theta }(\underline{x})$\ optimally
match $p(\underline{x})$. An appropriate objective function is the Kullback
information measure[6], otherwise known as the Minimum Cross-Entropy
principle[7]. Because this measure is asymmetric, we can apply it in two
different ways to this problem. Following[8,9,10] we call these the
"Cross-Entropy" and "Reverse Cross-Entropy" methods:%
\begin{eqnarray}
\text{CE} &\text{:}&\ \ \hat{q}_{\theta }\leftarrow \min\limits_{\underline{%
\theta }}H(q_{\theta },p) \\
\text{RCE} &\text{:}&\ \ \hat{q}_{\theta }\leftarrow \min\limits_{\underline{%
\theta }}H(p,q_{\theta })
\end{eqnarray}%
where:%
\begin{equation}
H(p_{1},p_{2})=\dint p_{1}(\underline{x})\log \frac{p_{1}(\underline{x})}{%
p_{2}(\underline{x})}d\underline{x}
\end{equation}%
Kullback[6] has argued that $H(p_{1},p_{2})$\ measures the mean amount of
information for discriminating in favor of the hypothesis that $p_{1}$ is
the correct density of \underline{$x$}\ rather than $p_{2}$. Shore and
Johnson[7] have argued that minimizing $H(p_{1},p_{2})$ over $p_{1}$ is the
only consistent estimation procedure for estimating a PDF given an a prior
density estimate $p_{2}(\underline{x})$\ combined with new structural
information about the density, such as one or more of its moments. The
measure $H(p_{1},p_{2})$ has several pleasing mathematical properties: it is
convex in $p_{1}$, and convex in $p_{2}$, and attains its minimum value of
zero when $p_{1}(\underline{x})=p_{2}(\underline{x})$\ almost everywhere.
Another useful property is that estimating \underline{$\theta $} from either
(3) or (4) is straightforward. Substitute (1) and (2) into the CE and RCE
formulas to obtain:%
\begin{eqnarray*}
\text{CE} &\text{:}&H(q_{\theta },p)=\xi \{tr(R^{-1}R_{\theta })-N-\log
\left\vert R^{-1}R_{\theta }\right\vert +(\underline{m}_{\theta }-\underline{%
m})^{H}R^{-1}(\underline{m}_{\theta }-\underline{m})\} \\
\text{RCE} &\text{:}&H(p,q_{\theta })=\xi \{tr(R_{\theta }^{-1}R)-N-\log
\left\vert R_{\theta }^{-1}R\right\vert +(\underline{m}_{\theta }-\underline{%
m})^{H}R_{\theta }^{-1}(\underline{m}_{\theta }-\underline{m})\}
\end{eqnarray*}%
where $\xi =1/2$ when \underline{$x$}\ is real and $\xi =1$ when \underline{$%
x$}\ is complex. \ 

To simplify the remainder of the discussion, assume that the mean is known, 
\underline{$m$}$_{\theta }=$\underline{$m$}, so that we can focus on the
estimation of the covariance matrix and compare the results with those by
Burg and Gray [4] and Gray, Anderson, Sim[5]. The two estimation problems
reduce to minimizing:%
\begin{eqnarray}
\text{CE} &\text{:}&\ \ \ \ H(q_{\theta },p)=\xi \left\{ tr(R^{-1}R_{\theta
})-N-\log \left\vert R^{-1}R_{\theta }\right\vert \right\} \\
\text{RCE} &\text{:}&\ \ \ \ H(p,q_{\theta })=\xi \{tr(R_{\theta
}^{-1}R)-N-\log \left\vert R_{\theta }^{-1}R\right\vert \}
\end{eqnarray}

Setting the gradients of the above two objective functions with respect to 
\underline{$\theta $}\ to zero, we obtain the necessary conditions that 
\underline{$\widehat{\theta }$} be the optimal solution:%
\begin{eqnarray}
\text{CE}\text{: } &&tr\left. \left\{ (R^{-1}-R_{\theta }^{-1})\frac{%
\partial R_{\theta }}{\partial \theta _{i}}\right\} \right\vert _{\underline{%
\theta }=\underline{\widehat{\theta }}}=0 \\
\text{RCE}\text{: } &&tr\left. \left\{ (R-R_{\theta })\frac{\partial
R_{\theta }^{-1}}{\partial \theta _{i}}\right\} \right\vert _{\underline{%
\theta }=\underline{\widehat{\theta }}}=0
\end{eqnarray}%
\bigskip for all $i$, where $\theta _{i}$ is the $i^{th}$ element of 
\underline{$\theta $}. When $R_{\theta }$\ is invertible and differentiable
in \underline{$\theta $}:%
\begin{equation}
\frac{\partial R_{\theta }^{-1}}{\partial \theta _{i}}=-R_{\theta }^{-1}%
\frac{\partial R_{\theta }}{\partial \theta _{i}}R_{\theta }^{-1}
\end{equation}%
Substituting this into the RCE formula gives an alternate set of necessary
conditions for the optimal RCE solution:%
\begin{equation}
\text{RCE:}\QTR{sl}{\ \ }tr\left. \left\{ (R_{\theta }^{-1}RR_{\theta
}^{-1}-R_{\theta }^{-1})\frac{\partial R_{\theta }}{\partial \theta _{i}}%
\right\} \right\vert _{\underline{\theta }=\underline{\widehat{\theta }}}=0
\end{equation}

\noindent \textbf{3. Application to Array Beamforming}

In this section we will apply the CE and RCE methods to fitting a low rank
plus noise covariance matrix to data. Such problems arise in a variety of
contexts, including narrowband sensor array processing and harmonic
retrieval. We focus on the former problem. Let \underline{$x$}$%
[n]=(x_{1}[n],...,x_{N}[n])^{T}$ be a vector of sensor measurements at time $%
n$, where $N$ is the total number of sensors in the array. Assume that the
signal is narrowband (perhaps because the sensor data has been preprocessed
through a Fast Fourier Transform of each sensor's data). Let our initial PDF
estimate for the data be given by $p(\underline{x}[n])=N(\underline{0},R),$
where $R$ is any non-parameterized estimate of the signal covariance, such
as $R=\frac{1}{K}\sum_{k=1}^{K}\underline{x}\left[ k\right] \underline{x}^{H}%
\left[ k\right] $ where $K$ snapshots of array data are used.

Now suppose we wish to model the data $\underline{x}[n]$ as: 
\begin{equation}
\underline{x}[n]=\sum_{i=1}^{P}s_{i}[n]\underline{u}_{i}+\sigma \underline{w}%
[n]
\end{equation}%
\ \ \ \ \ \ \ \ \ \ \ \ \ \ \ \ \ \ \ \ \ \ \ \ \ \ \ \ \ \ \ \ \ \ \ \ \ \
where $s_{1}[n],....,s_{P}[n]$\ are $P$ source signals, $P<N$, arriving from
unknown directions \underline{$u$}$_{1},...,\underline{u}_{P},$ with
additive noise \underline{$w$}$[n]$ with gain $\sigma $. Suppose that
signals $s_{i}[n]$\ are statistically independent, real or complex zero mean
Gaussian random variables with covariance $\Lambda _{i}>0$, and that the
noise samples \underline{$w$}$[n]$ are statistically independent, real or
complex zero mean Gaussian random variables with covariance $W$.%
\begin{eqnarray}
p(s_{i}[n]) &=&N(0,\Lambda _{i}) \\
p(\underline{w}[n]) &=&N(\underline{0},W)
\end{eqnarray}

Thus the parameterized model PDF of \underline{$x$}$[n]$ is Gaussian:%
\begin{equation}
q_{\theta }(\underline{x}[n])=N(\underline{0},R_{\theta })
\end{equation}%
where:%
\begin{equation}
R_{\theta }=\sum_{i=1}^{P}\Lambda _{i}\underline{u}_{i}\underline{u}%
_{i}^{H}+\sigma ^{2}W
\end{equation}%
We will assume that the noise covariance $W$ is known, but that all the
other parameters \underline{$\theta $}$=(\Lambda _{1},...,\Lambda _{P},%
\underline{u}_{1},...\underline{u}_{P},\sigma )^{T}$ must be estimated. For
convenience, define:%
\begin{equation}
R_{\theta }=U\Lambda U^{H}+\sigma ^{2}W
\end{equation}%
where:%
\begin{equation}
U=\left[ 
\begin{array}{cccc}
\underline{u}_{1} & \underline{u}_{2} & ... & \underline{u}_{P}%
\end{array}%
\right] \text{ and\textsl{\ }}\Lambda =%
\begin{bmatrix}
\Lambda _{1} &  & 0 \\ 
& \ddots &  \\ 
0 &  & \Lambda _{P}%
\end{bmatrix}%
\end{equation}%
Suppose there are no a priori constraints on the matrix $U$, and that the
only constraints on $\Lambda $\ are that $\Lambda _{i}>0$. This would
typically be true if the array were uncalibrated, or subject to heavy
unknown multipath distortion. (Note that because we assume an uncalibrated
array, we will not be able to directly derive information about the
direction of arrival.) Appendices A and B apply the CE and RCE criteria to
this model. They show that the solution to these two problems are quite
similar, and can be found by the following algorithm:

\vspace{0.2in} {\raggedright\textbf{CE and RCE BEAMFORMING ALGORITHMS}}

\begin{enumerate}
\item Find the generalized eigenvector \underline{$u$}$_{i}$ and eigenvalue $%
\lambda _{i}$ solutions to:%
\begin{equation}
\lambda _{i}R^{-1}\underline{u}_{i}=W^{-1}\underline{u}_{i}
\end{equation}%
with normalization constraint \underline{$u$}$_{i}^{H}W^{-1}$\underline{$u$}$%
_{j}=\delta _{i,j}$.

\item Sort the eigenvectors and eigenvalues so that $\lambda _{1}\geq
\lambda _{2}\geq ...\geq \lambda _{N}.$ Then the optimal structured
covariance matrix approximation $\hat{R}_{\theta }$ to $R$ is :%
\begin{equation}
\hat{R}_{\theta }=\left( 
\begin{array}{cccc}
\underline{u}_{1} & \underline{u}_{2} & ... & \underline{u}_{P}%
\end{array}%
\right) 
\begin{pmatrix}
\lambda _{1}-\widehat{\sigma }^{2} &  & 0 \\ 
& \ddots &  \\ 
0 &  & \lambda _{P}-\widehat{\sigma }^{2}%
\end{pmatrix}%
\begin{pmatrix}
\underline{u}_{1}^{H} \\ 
\underline{u}_{2}^{H} \\ 
. \\ 
. \\ 
. \\ 
\underline{u}_{P}^{H}%
\end{pmatrix}%
+\widehat{\sigma }^{2}W
\end{equation}%
where: 
\begin{equation}
\left\{ 
\begin{array}{c}
\frac{1}{\widehat{\sigma }^{2}}=\frac{1}{N-P}\sum\limits_{i=P+1}^{N}\frac{1}{%
\lambda _{i}}\text{ \ for CE} \\ 
\widehat{\sigma }^{2}=\frac{1}{N-P}\sum\limits_{i=P+1}^{N}\lambda _{i}\text{
\ for RCE}%
\end{array}%
\right.
\end{equation}

\item The cross entropy for the optimal model is:%
\begin{eqnarray}
\text{CE}\text{: \ \ \ } &&H(\hat{q}_{\theta },p)=\xi \sum_{i=P+1}^{N}\log (%
\frac{\lambda _{i}}{\widehat{\sigma }^{2}}) \\
\text{RCE}\text{: \ \ } &&H(p,\hat{q}_{\theta })=\xi \sum_{i=P+1}^{N}\log (%
\frac{\hat{\sigma}^{2}}{\lambda _{i}})
\end{eqnarray}
\end{enumerate}

The estimates of $\hat{R}_{\theta }$ \ and $\hat{\sigma}^{2}$\ will be
unique if and only if $\lambda _{P}>\lambda _{P+1}.$ (The estimate of $U$
will not be unique.)

An interesting alternative form for the cross-entropy formulas can be found
by substituting the value of $\hat{\sigma}^{2}$\ from (21) into (22):%
\begin{eqnarray}
\text{CE}\text{: \ } &&H(\hat{q}_{\theta },p)=\xi (N-P)\log \left( \frac{%
\left[ \frac{1}{\lambda _{P+1}},...,\frac{1}{\lambda _{N}}\right] _{avg}}{%
\left[ \frac{1}{\lambda _{P+1}},...,\frac{1}{\lambda _{N}}\right] _{geo}}%
\right) \\
\text{RCE}\text{: \ } &&H(p,\hat{q}_{\theta })=\xi (N-P)\log \left( \frac{%
\left[ \lambda _{P+1},...,\lambda _{N}\right] _{avg}}{\left[ \lambda
_{P+1},...,\lambda _{N}\right] _{geo}}\right)
\end{eqnarray}%
where:%
\begin{eqnarray}
\left[ \beta _{P+1},...,\beta _{N}\right] _{avg} &=&\frac{1}{N-P}%
\sum_{i=P+1}^{N}\beta _{i} \\
\lbrack \beta _{P+1},...,\beta _{N}]_{geo} &=&(\beta _{P+1}\beta
_{P+2}...\beta _{N})^{1/(N-P)}
\end{eqnarray}%
The cross-entropies are proportional to the log of the ratio of the
arithmetic mean to the geometric mean of the eigenvalues (or their inverses)
that are not used in building $U$. The cross-entropy will therefore be
positive, and will attain their minimum value of zero only if the geometric
average of $\lambda _{P+1},...,\lambda _{N}$ (or their inverses) equals
their arithmetic mean. This will only occur if these $N-P$ smallest
generalized eigenvalues are all equal.

Note the similarity of the RCE formula to the MDL order determination
algorithm suggested by Wax and Kailath[2]. The RCE criterion is also
strongly related to the Maximum Likelihood problem of estimating the
structured covariance matrix given observations \underline{$x$}$_{1},...,$%
\underline{$x$}$_{K}$: 
\begin{equation}
\hat{R}_{\theta }\leftarrow \max_{\theta }\log p(\underline{x}_{1},...,%
\underline{x}_{K}\mid \underline{\theta })
\end{equation}%
where:%
\begin{equation}
p(\underline{x}_{1},...,\underline{x}_{K}\mid \underline{\theta })\text{ }%
=\dprod\limits_{i=1}^{K}p(\underline{x}_{i}\mid \underline{\theta })
\end{equation}%
and:%
\begin{equation}
p(\underline{x}_{i}\mid \underline{\theta })=N(\underline{0},R)
\end{equation}%
This is because:%
\begin{equation}
H(p,q_{\theta })=\frac{1}{K}\log p(\underline{x}_{1},...,\underline{x}%
_{K}\mid \underline{\theta })-\xi (N+\log \left\vert R\right\vert )
\end{equation}%
\bigskip Since the second term in (31) does not depend on \underline{$\theta 
$}, the RCE estimate of $R_{\theta }$\ will be identical to the ML estimate.

For the special case when the background noise is white Gaussian noise, $W=I$%
, the \underline{$u$}$_{i}$ must satisfy:%
\begin{equation}
R\underline{u}_{i}=\lambda _{i}\underline{u}_{i}
\end{equation}%
and thus the \underline{$u$}$_{i\text{\textsl{\ }}}$are the eigenvectors of
the observed data correlation matrix $R$. This special case is thus quite
similar to that used in the MUSIC algorithm[1] and other similar beamforming
algorithms.

If subroutines for computing generalized eigenvectors are not available, we
can use subroutines for computing eigenvectors of symmetric positive
definite matrices as follows. Factor $W=W^{1/2}W^{H/2}$\ where $W^{1/2}$\ is
any square root of $W$ and $W^{H/2}$\ is its Hermitian. Then to compute the 
\underline{$u$}$_{i}$:

\begin{enumerate}
\item From the whitened data correlation matrix:%
\begin{equation}
\tilde{R}=W^{-1/2}RW^{-H/2}
\end{equation}%
where $W^{-1/2}\ $is the inverse of $W^{1/2}.$ Note that $\tilde{R}$ is
symmetric and positive definite.

\item Solve for the eigenvectors \underline{$t$}$_{i}$ and corresponding
eigenvalues $\lambda _{i}$\ of $\tilde{R}.$%
\begin{equation}
\tilde{R}\underline{t}_{i}=\lambda _{i}\underline{t}_{i}
\end{equation}%
where \underline{$t$}$_{j}^{T}\underline{t}_{i}=\delta _{i.j}$. Sort these
so that the eigenvalues are in descending order.

\item Then:%
\begin{equation}
\underline{u}_{i}=W^{1/2}\underline{t}_{i}
\end{equation}
\end{enumerate}

It is also interesting to consider the effect of using the structured
covariance matrix estimate when forming either a classical or optimal
beamformer. Let \underline{$w$}$_{0}\ $be the ideal array response for a
signal in a particular direction. The classical beamformer estimates the
signal $s[n]$\ from the array data as $s[n]=\underline{w}_{0}^{T}\underline{x%
}[n]$. The expected received power from this direction is then $E[s^{2}[n]]=%
\underline{w}_{0}^{T}R_{\theta }\underline{w}_{0}.$\ Now suppose that 
\underline{$w$}$_{o}$\ is in the space spanned\ by the columns of $R^{-1}U$,
i.e. \underline{$w$}$_{0}=R^{-1}U$\underline{$\alpha $} for some vector 
\underline{$\alpha $}. It is shown in Appendix A that $R_{\theta
}^{-1}U=R^{-1}U$. Therefore:%
\begin{eqnarray}
\underline{w}_{0}^{H}R_{\theta }\underline{w}_{0} &=&\underline{\alpha }%
^{H}U^{H}R^{-H}R_{\theta }R^{-1}U\underline{\alpha }  \notag \\
&=&\underline{\alpha }^{H}U^{H}R^{-H}U\underline{\alpha }  \notag \\
&=&\underline{w}_{0}^{H}R\underline{w}_{0}
\end{eqnarray}%
In this case, replacing $R$ with the structured covariance estimate $%
R_{\theta }$ in the classical beamformer makes no difference. However, if $%
\underline{w}_{0}$ is not in the subspace spanned by $R^{-1}\underline{u}%
_{1},...,R^{-1}\underline{u}_{P}$, then $R_{\theta }^{-1}\underline{w}%
_{0}\neq R^{-1}\underline{w}_{0}$, and using the structured covariance
estimate in the classical beamformer will yield a different beam pattern.

A similar statement holds for the optimum minimum variance beamformer, $s[n]=%
\underline{w}^{T}\underline{x}[n],$ which uses a window \underline{$w$}
designed such that the expected response energy \underline{$w$}$%
^{T}R_{\theta }$\underline{$w$}\ is minimized subject to the constraint that
the response to a plane wave from the direction of interest is unity, 
\underline{$w$}$^{T}\underline{w}_{0}=1$. The solution is $\underline{w}%
=\left( \underline{w}_{0}^{T}R_{\theta }^{-1}\underline{w}_{0}\right)
^{-1}R_{\theta }^{-1}\underline{w}_{0}$. Note that if \underline{$w$}$_{0}$\
is in the subspace spanned by the columns of \ $U$, then there exists some
vector \underline{$\alpha $}\ such that \underline{$w$}$_{0}=U$\underline{$%
\alpha $}. Since $R_{\theta }^{-1}U=R^{-1}U,$%
\begin{equation}
R_{\theta }^{-1}\underline{w}_{0}=R_{\theta }^{-1}U\underline{\alpha }%
=R^{-1}U\underline{\alpha }=R^{-1}\underline{w}_{0}
\end{equation}%
which in turn implies:%
\begin{equation}
\underline{w}=\left( \underline{w}_{0}^{T}R_{\theta }^{-1}\underline{w}%
_{0}\right) ^{-1}R_{\theta }^{-1}\underline{w}_{0}=\left( \underline{w}%
_{0}^{T}R^{-1}\underline{w}_{0}\right) ^{-1}R^{-1}\underline{w}_{0}
\end{equation}%
In this case, replacing $R$ with the structured covariance estimate $%
R_{\theta }$\ in the optimal beamformer makes no difference. However, if 
\underline{$w$}$_{0}$ is not in the subspace spanned by the columns of $U$,
then $R_{\theta }^{-1}\underline{w}_{0}\neq R^{-1}\underline{w}_{0}$, and
using the structured covariance estimate in the optimal beamformer will
yield a different beam pattern. These results are contrary to the suggestion
implied in [5] that replacing $R$ with $R_{\theta }$\ in an optimal
beamformer should make no difference.

\bigskip

\noindent \textbf{4. Conclusion}

\bigskip In this paper, we have derived the optimal solution for correlation
matrix estimation by the CE and RCE principles. The two methods give
identical results in the problem of estimating the sum of a low rank signal
matrix plus noise matrix, differing only in the value of the noise level
estimate. The RCE method gives the same results as the Maximum Likelihood
approach, and when the noise is white, both methods are similar to MUSIC. It
is interesting that the cross-entropy approach thus provides a unifying
framework for deriving spectral estimation algorithm including Bartlett,
MLM[8], MEM[10], and now MUSIC.

\bigskip \newpage

\noindent {\Large A \ \ Derivation of CE Beamforming Algorithm}

\bigskip In this appendix we derive the optimal structured covariance
estimate using the CE principle. First, to simplify the effort, let us
define: $V=U\Lambda ^{1/2}$, where $\Lambda ^{1/2}=diag(\Lambda
_{1}^{1/2},...,\Lambda _{N}^{1/2})$. Then:%
\begin{equation}
R_{\theta }=VV^{H}+\sigma ^{2}W
\end{equation}%
Substitute this into the CE entropy expression (6), and set the derivatives
with respect to the real and imaginary part of every element of the $V$
matrix, and with respect to $\sigma ^{2}$, to zero. Arranging these
derivatives in complex matrix form gives:%
\begin{eqnarray}
(R^{-1}-R_{\theta }^{-1})V &=&0 \\
tr\{(R^{-1}-R_{\theta }^{-1})W\} &=&0
\end{eqnarray}%
Using the Woodward lemma:%
\begin{equation}
R_{\theta }^{-1}=\frac{1}{\sigma ^{2}}W^{-1}-\frac{1}{\sigma ^{2}}W^{-1}V%
\text{ }\left[ V^{H}\frac{1}{\sigma ^{2}}W^{-1}V+I\right] ^{-1}V^{H}\frac{1}{%
\sigma ^{2}}W^{-1}
\end{equation}%
Substituting into (40) and simplifying gives:%
\begin{equation}
R^{-1}V=\frac{1}{\sigma ^{2}}W^{-1}V\left[ V^{H}\frac{1}{\sigma ^{2}}%
W^{-1}V+I\right] ^{-1}
\end{equation}%
This equation has many possible solutions. Let $V$ refer to any one of
these. Then let $\Psi =V^{H}W^{-1}V$. Diagonalize $\Psi $ by factoring it: $%
\Psi =Q\Phi Q^{H}$, where $\Phi $\ is diagonal and $Q$ is orthonormal, $%
Q^{H}Q=I$. Define $\tilde{V}=VQ$. Note that $\tilde{V}$\ is also a solution
to (43). In fact,%
\begin{equation}
R^{-1}\tilde{V}=\frac{1}{\sigma ^{2}}W^{-1}\tilde{V}\left[ \frac{1}{\sigma
^{2}}\Phi +I\right] ^{-1}
\end{equation}%
and:%
\begin{equation}
\tilde{V}^{H}W^{-1}\tilde{V}=\Phi
\end{equation}%
Let the $P$ columns of $\tilde{V}$ be \underline{$\tilde{v}$}$_{1,}...,%
\underline{\tilde{v}}_{P},$ and let the $P$ diagonal elements of $\Phi $\ be 
$\phi _{1},...,\phi _{P}$. Then:%
\begin{equation}
\lambda _{i}R^{-1}\underline{\tilde{v}}_{i}=W^{-1}\underline{\tilde{v}}_{i}
\end{equation}%
where:%
\begin{equation}
\lambda _{i}=\phi _{i}+\sigma ^{2}
\end{equation}%
The columns of $\tilde{V}$ must therefore either be zero, or else must be
generalized eigenvector solutions to (46). Because $R$ and $W$ are conjugate
symmetric and positive definite, there are always $N$ linearly independent
generalized eigenvector solutions \underline{$\tilde{v}$}$_{1},...,%
\underline{\tilde{v}}_{N}$ to (46), with corresponding generalized
eigenvalues $\lambda _{1},...,\lambda _{N}$ which are positive. Assume
without loss of generality that the first $P_{0}$\ columns of $\tilde{V}$
are non-zero, where $P_{0}\leq P$\ . These first $P_{0}$\ columns must be
selected from among the $N$ possible generalized eigenvectors, in a manner
we will determine later. Also note that it is not necessary to estimate $Q$
or $V$ directly, since we can construct $R_{\theta }$\ directly from $\tilde{%
V}$:%
\begin{eqnarray}
R_{\theta } &=&VV^{H}+\sigma ^{2}W  \notag \\
&=&VQQ^{H}V^{H}+\sigma ^{2}W  \notag \\
&=&\tilde{V}\tilde{V}^{H}+\sigma ^{2}W
\end{eqnarray}

Now to solve for $\sigma ^{2}$. Substitute (42) into (41), and simplify by
exploiting the facts that $tr(AB)=tr(BA)$\ and $tr(C+D)=tr(C)+tr(D)$\ and $%
tr(\alpha C)=\alpha tr(C)$\ where $A,B$ are matrices, $C,D$ are square
matrices, and $\alpha $\ is a scalar.

\bigskip

\begin{eqnarray}
0 &=&tr\{(R_{\theta }^{-1}-R^{-1})W\}  \notag \\
&=&tr\left\{ \left( \frac{1}{\sigma ^{2}}W^{-1}-\frac{1}{\sigma ^{2}}W^{-1}%
\tilde{V}\left[ \tilde{V}^{H}\frac{1}{\sigma ^{2}}W^{-1}\tilde{V}+I\right]
^{-1}\tilde{V}^{H}\frac{1}{\sigma ^{2}}W^{-1}-R^{-1}\right) W\right\}  \notag
\\
&=&tr\left\{ \frac{1}{\sigma ^{2}}I\right\} -\frac{1}{\sigma ^{2}}tr\left\{ %
\left[ \tilde{V}^{H}\frac{1}{\sigma ^{2}}W^{-1}\tilde{V}+I\right] ^{-1}\left[
\tilde{V}^{H}\frac{1}{\sigma ^{2}}W^{-1}\tilde{V}\right] \right\} -tr\left\{
R^{-1}W\right\}  \notag \\
&=&\frac{N}{\sigma ^{2}}-\frac{1}{\sigma ^{2}}\sum_{i=1}^{P}\frac{\phi _{i}}{%
\phi _{i}+\sigma ^{2}}-tr\{R^{-1}W\}  \notag \\
&=&\frac{N-P_{0}}{\sigma ^{2}}+\sum_{i=1}^{P_{0}}\frac{1}{\lambda _{i}}%
-tr\{WR^{-1}\}\text{ }
\end{eqnarray}%
where we used (45) in the fourth line, and (47) in the fifth. This can be
further simplified by noticing that if \underline{$\tilde{v}$}$_{i}$\ is any
generalized eigenvector solution to (46), then:%
\begin{equation}
WR^{-1}\underline{\tilde{v}}_{i}=W(\frac{1}{\lambda _{i}}W^{-1}\underline{%
\tilde{v}}_{i})=\frac{1}{\lambda _{i}}\underline{\tilde{v}}_{i}
\end{equation}%
Therefore, the \underline{$\tilde{v}$}$_{i}$ are eigenvectors of $WR^{-1}$\
with eigenvalues $1/\lambda _{i}$. Thus:%
\begin{equation}
tr\{WR^{-1}\}=\sum_{i=1}^{N}\frac{1}{\lambda _{i}}
\end{equation}%
Substituting back into (49), then solving for $\sigma ^{2}$\ gives:%
\begin{equation}
\sigma ^{2}=\frac{N-P_{0}}{\sum\limits_{i=P_{0}+1}^{N}\frac{1}{\lambda _{i}}}
\end{equation}

Now substitute the solution for $\tilde{V}$\ and for $\sigma ^{2}$\ into
(48), and then substitute this back into the formula (6) for the
cross-entropy. The algebra is simplified by noting that if \underline{$%
\tilde{v}$}$_{i}$\ is any generalized eigenvector solution to (46), then:%
\begin{eqnarray}
R_{\theta }R^{-1}\underline{\tilde{v}}_{i} &=&(\tilde{V}\tilde{V}^{H}+\sigma
^{2}W)\left( \frac{1}{\lambda _{i}}W^{-1}\underline{\tilde{v}}_{i}\right) 
\notag \\
&=&\frac{1}{\lambda _{i}}(\tilde{V}\tilde{V}^{H}W^{-1}\underline{\tilde{v}}%
_{i}+\sigma ^{2}\underline{\tilde{v}}_{i})  \notag \\
&=&\left\{ 
\begin{array}{cl}
\frac{1}{\lambda _{i}}(\phi _{i}+\sigma ^{2})\underline{\tilde{v}}_{i} & 
\text{for }i=1,....,P_{0} \\ 
\frac{1}{\lambda _{i}}\sigma ^{2}\underline{\tilde{v}}_{i} & \text{for }i%
\text{ }=P_{0}+1,...,N%
\end{array}%
\right.  \notag \\
&=&\left\{ 
\begin{array}{cl}
\underline{\tilde{v}}_{i} & \text{for }i=1,....,P_{0} \\ 
\frac{\sigma ^{2}}{\lambda _{i}}\underline{\tilde{v}}_{i} & \text{for }%
i=P_{0}+1,...,N%
\end{array}%
\right.
\end{eqnarray}%
Therefore, the \underline{$\tilde{v}$}$_{i}$ are all eigenvectors of $%
R_{\theta }R^{-1}$. The first $P_{0}\ $eigenvalues are equal to 1, and the
remainder are equal to $\sigma ^{2}/\lambda _{P_{0}+1},...,\sigma
^{2}/\lambda _{N}$. Putting all this together, the cross-entropy at this
solution has the value:%
\begin{eqnarray}
H(q_{\theta },p) &=&\xi \left\{ tr\{R_{\theta }R^{-1}\}-N-\log \left\vert
R_{\theta }R^{-1}\right\vert \right\}  \notag \\
&=&\xi \left\{ P_{0}+\sigma ^{2}\sum_{i=P_{0}+1}^{N}\frac{1}{\lambda _{i}}%
-N-\log \prod_{i=P_{0}+1}^{N}\text{\ \ }\frac{\sigma ^{2}}{\lambda _{i}}%
\right\}  \notag \\
&=&\xi \sum_{i=P_{0}+1}^{N}\log \left( \frac{\lambda _{i}}{\sigma ^{2}}%
\right)
\end{eqnarray}%
Substituting the value of $\sigma ^{2}$\ from (52) gives the alternate form:%
\begin{equation}
H(q_{\theta },p)=\xi (N-P_{0})\log \left[ \frac{\frac{1}{N-P_{0}}%
\sum\limits_{i=P_{0}+1}^{N}\frac{1}{\lambda _{i}}}{\left(
\dprod\limits_{i=P_{0}+1}^{N}\frac{1}{\lambda _{i}}\right) ^{1/(N-P_{0})}}%
\right]
\end{equation}

Now to return to the issue of which of the $N$ possible generalized
eigenvector solutions should be used for the $P_{0}\ $non-zero columns of $%
\tilde{V}$. Let us call the selected $P_{0}$ eigenvectors $\underline{\tilde{%
v}}_{1},...,\underline{\tilde{v}}_{P_{0}}$\ the "signal eigenvectors", and
let us call the remainder the "noise eigenvectors". The signal eigenvectors
satisfy $\underline{\tilde{v}}_{i}\neq 0$; since $W^{-1}>0$, then $\phi _{i}=%
\underline{\tilde{v}}_{i}^{H}W^{-1}\underline{\tilde{v}}_{i}>0$\ and thus $%
\lambda _{i}=\phi _{i}+\sigma ^{2}>\sigma ^{2}$ \ for $i=1,...,P_{0}$. We
show that these signal eigenvalues must be the largest eigenvalue solutions
to (46). Suppose this were not true, so that the global optimum solution
corresponded to an $R_{\theta }$ such that one of the signal eigenvalues,
say $\lambda _{P_{0}},$ was smaller than the largest of the noise
eigenvalues, say $\lambda _{P_{0}+1}$. Thus $\sigma ^{2}<\lambda
_{P_{0}}<\lambda _{P_{0}+1}$. But then, as we will see, swapping these
eigensolutions, making $\underline{\tilde{v}}_{P_{0}+1}$ a signal
eigenvector and $\underline{\tilde{v}}_{P_{0}}$ a noise eigenvector will
further decrease the cross-entropy, contradicting our assumption of global
optimality. To show this, let $H\left( \lambda _{P_{0}+1},\lambda
_{P_{0}+2},...,\lambda _{N}\right) $ represent the cross-entropy with a
model $R_{\theta }$ built using non-zero solutions $\underline{\tilde{v}}%
_{1},...,\underline{\tilde{v}}_{P_{0}-1},\underline{\tilde{v}}_{P_{0}}$, and
let $H\left( \lambda _{P_{0}},\lambda _{P_{0}+2},...,\lambda _{N}\right) $
represent the cross-entropy with a model $R_{\theta }$ built using non-zero
solutions $\underline{\tilde{v}}_{1},...,\underline{\tilde{v}}_{P_{0}-1},%
\underline{\tilde{v}}_{P_{0}+1}$. Then because the cross-entropy formula
(55) is an analytic function of the $\lambda _{i}$, by the mean value
theorem:%
\begin{eqnarray}
&&H(\lambda _{P_{0}+1,}\lambda _{P_{0}+2,}...,\lambda _{N})-H(\lambda
_{P_{0},}\lambda _{P_{0}+2,}...,\lambda _{N})  \notag \\
&=&\frac{\partial H}{\partial \lambda }(\lambda ,\lambda
_{P_{0}+2,}....,\lambda _{N}){\Huge \mid }_{\lambda =\bar{\lambda}}(\lambda
_{P_{0}+1}-\lambda _{P_{0}})
\end{eqnarray}%
where $\bar{\lambda}$\ is some value in the range $\lambda _{P_{0}}<\bar{%
\lambda}<\lambda _{P_{0}+1}$. But:%
\begin{eqnarray}
\frac{\partial H}{\partial \lambda } &=&\xi \frac{1}{\lambda ^{2}}\left(
\lambda -\frac{N-P_{0}}{\frac{1}{\lambda }+\sum\limits_{i=P_{0}+2}^{N}\frac{1%
}{\lambda _{i}}}\right)  \notag \\
&>&0
\end{eqnarray}%
for all $\lambda _{P_{0}}<\lambda <\lambda _{P_{0}+1}$, where the last line
is true because:%
\begin{eqnarray}
\lambda &>&\lambda _{P_{0}}  \notag \\
&>&\sigma ^{2}  \notag \\
&=&\frac{N-P_{0}}{\sum_{i=P_{0}+1}^{N}\frac{1}{\lambda _{i}}}  \notag \\
&>&\frac{N-P_{0}}{\frac{1}{\lambda }+\sum_{i=P_{0}+2}^{N}\frac{1}{\lambda
_{i}}}
\end{eqnarray}%
Since $\lambda _{P_{0}+1}-\lambda _{P_{0}}>0$, the change in (56) must be
positive. Therefore, swapping $\underline{\tilde{v}}_{P_{0}}$ and $%
\underline{\tilde{v}}_{P_{0}+1}$\ reduces the cross-entropy, and our assumed
global optimum solution cannot be globally optimum. The $P_{0}$\ signal
eigenvalues must therefore be the largest eigenvalue solutions to (46), and
the non-zero $P_{0}$\ columns of $\tilde{V}$\ must be the corresponding
general eigenvectors.

Finally, we must show that we should always choose $P_{0}=P$\ eigenvectors.
Without loss of generality, let us sort all the eigenvalues $\lambda
_{1}\geq \lambda _{2}\geq ...\geq \lambda _{N}$. Let $H_{i}$\ represent the
minimum cross-entropy with $i$ non-zero columns in $\tilde{V}$. Then using
(55):%
\begin{eqnarray}
H_{P_{0}}-H_{P_{0}+1} &=&\xi (N-P_{0})\log \left[ \frac{\frac{1}{(N-P_{0})}%
\frac{1}{\lambda _{P_{0}+1}}+\left( 1-\frac{1}{(N-P_{0})}\right) \frac{1}{%
\bar{\lambda}}}{\left( \frac{1}{\lambda _{P_{0}+1}}\right) ^{\frac{1}{\left(
N-P_{0}\right) }}\left( \frac{1}{\bar{\lambda}}\right) ^{1-\frac{1}{(N-P_{0})%
}}}\right]  \notag \\
&\geq &0
\end{eqnarray}%
where $\frac{1}{\bar{\lambda}}=\frac{1}{N-P_{0}-1}$\ $\sum_{i=P_{0}+2}^{N}%
\frac{1}{\lambda _{i}}$ and where we used the inequality $\rho \alpha
+(1-\rho )\beta \geq \alpha ^{\rho }\beta ^{(1-\rho )}$\ for any $0\leq \rho
\leq 1$\ in the last line. Thus the cross-entropy decreases as $P_{0}$\
varies from $0$\ to $P$, so the best choice for $P_{0}$\ must be $P_{0}=P$\ .

The proof that $R_{\theta }$\ is unique when $\lambda _{P}>\lambda _{P+1}$\
is messy but straightforward. The key issue is that the space spanned by the
signal eigenvectors is uniquely determined. If there are multiple signal
eigenvalues, then the eigenvectors themselves may not be uniquely
determined, and thus $\tilde{V}$\ may not be uniquely determined.

We get the formulas in the text by defining $U=\tilde{V}\Phi ^{-1/2}$\ .

\bigskip \newpage

\noindent {\Large B \ \ \ Derivation of RCE Algorithm}

In this appendix we give the solution to the RCE problem. The derivation is
quite similar to that for the CE problem, and therefore we present this
quickly. With our Gaussian models, the RCE cross-entropy has the value:%
\begin{equation}
\text{RCE: \ }H(p,q_{\theta })=\xi \left\{ tr(R_{\theta }^{-1}R)-N-\log
\left\vert R_{\theta }^{-1}R\right\vert \right\}
\end{equation}%
Differentiating with respect to the real and imaginary parts of $V$\ and
setting these to zero, as before, gives:%
\begin{equation}
\left( R_{\theta }^{-1}RR_{\theta }^{-1}-R_{\theta }^{-1}\right) V=0
\end{equation}%
Multiplying both sides by $R^{-1}R_{\theta }$\ gives:%
\begin{equation}
(R_{\theta }^{-1}-R^{-1})V=0
\end{equation}%
which is exactly the same equations which the solution for $V$\ in the CE
problem must satisfy, (40). Therefore, we can construct $R_{\theta }\ $from
(48), where the columns of $\tilde{V}$\ must be solutions to the generalized
eigenvector problem (46).

Now differentiating (60) with respect to $\sigma ^{2}$ and setting it to
zero gives:%
\begin{equation}
tr\{(R_{\theta }^{-1}RR_{\theta }^{-1}-R_{\theta }^{-1})W\}=0
\end{equation}%
Combining this with (61) gives:%
\begin{eqnarray}
0 &=&tr\{(R_{\theta }^{-1}RR_{\theta }^{-1}-R_{\theta }^{-1})(\tilde{V}%
\tilde{V}^{H}+\sigma ^{2}W)\}  \notag \\
&=&tr\{(R_{\theta }^{-1}RR_{\theta }^{-1}-R_{\theta }^{-1})R_{\theta }\} 
\notag \\
&=&tr\{R_{\theta }^{-1}R-I\}
\end{eqnarray}%
which implies that:%
\begin{equation}
tr\{R_{\theta }^{-1}R\}=N\text{ }
\end{equation}%
But (53) implies that $R_{\theta }^{-1}R\ $has $P_{0}$ eigenvalues equal to $%
1$, and the rest have values $\lambda _{P_{0}+1}/\sigma ^{2},...,\lambda
_{N}/\sigma ^{2}$. Since the trace of a matrix is just the sum of its
eigenvalues:%
\begin{equation}
P_{0}+\frac{1}{\sigma ^{2}}\sum_{i=P_{0}+1}^{N}\lambda _{i}=N
\end{equation}%
which gives:%
\begin{equation}
\sigma ^{2}=\frac{1}{(N-P_{0})}\sum_{i=P_{0}+1}^{N}\lambda _{i}
\end{equation}%
Using the facts that the trace of a matrix is the sum of the eigenvalues,
and the determinant is the product of the eigenvalues:%
\begin{eqnarray}
H(p,q_{\theta }) &=&\xi \{tr\{R_{\theta }^{-1}R\}-N-\log \left\vert
R_{\theta }^{-1}R\right\vert \}  \notag \\
&=&\xi \sum_{i=P_{0}+1}^{N}\log \frac{\sigma ^{2}}{\lambda _{i}}
\end{eqnarray}%
The proofs that we must choose $\lambda _{1},...,\lambda _{P_{0}}$\ to be
the largest eigenvalues, that we should choose $P_{0}=P$\ , and that the
solution $R_{\theta }$\ is unique if $\lambda _{P}>\lambda _{P+1}$, are
similar to the proofs for the CE algorithm.

\bigskip

\bigskip

\noindent $^{1}$Department of Computer Science and Information Engineering,
National Taiwan University, Taipei, Taiwan, 106, Republic of China, Tel.:886
2 23625336 ext.515, Fax.:886 2 23628167, \noindent Email:
cyliou@csie.ntu.edu.tw

\noindent $^{2}$was with Massachusetts Institute of Technology, Research
Laboratory of Electronics, Cambridge, MA 02139.

\bigskip

\end{document}